\def\countMode{0}

\def\hlMode{0}

\documentclass[%
 aip,
 amsmath,amssymb,
 reprint,%
]{revtex4-1}

\usepackage{graphicx}
\usepackage{dcolumn}
\usepackage{bm}
\usepackage{color,soul}

\usepackage[utf8]{inputenc}
\usepackage[T1]{fontenc}
\usepackage{mathptmx}
\usepackage{etoolbox}

\newcommand{\U}[1]{\mathrm{\; #1}}
\newcommand{\E}[1]{\cdot 10^{#1}}
\newcommand{\Eb}[1]{10^{#1}}

\newcommand{\lb}{\left(}
\newcommand{\rb}{\right)}
\newcommand{\br}[1]{\lb #1 \rb} 
\newcommand{\sbr}[1]{\left[ #1 \right]} 
\newcommand{\pbr}[1]{\left\{ #1 \right\}} 

\renewcommand{\vec}[1]{\bm{#1}}
\newcommand{\vB}{\vec{B}}
\newcommand{\vE}{\vec{E}}

\newcommand{\vj}{\vec{j}}
\newcommand{\vv}{\vec{v}}

\newcommand{\uniVec}[1]{\vec{\hat{#1}}}
\newcommand{\uniR}{\uniVec{r}}
\newcommand{\uniPhi}{\uniVec{\phi}}
\newcommand{\uniZ}{\uniVec{z}}

\newcommand{\grad}{\vec{\nabla}}

\renewcommand{\div}{\vec{\nabla \cdot}}
\newcommand{\matDeriv}[1]{\br{#1 \vec{\cdot \nabla}}} 
\newcommand{\matDerivSq}[1]{\sbr{#1 \vec{\cdot \nabla}}} 
\newcommand{\curl}{\vec{\nabla \times}}
\newcommand{\bcurl}[1]{\curl \sbr{#1}}  

\newcommand{\crp}[2]{#1 \vec{\times} #2} 
\newcommand{\bcrp}[2]{\crp{#1}{\br{#2}}} 
\newcommand{\bcrpSq}[2]{\crp{#1}{\sbr{#2}}} 
\newcommand{\dtp}[2]{#1 \vec{\cdot} #2}  
\newcommand{\curlB}{\curl \vB}           
\newcommand{\bcurlB}{\br{\curlB}}        
\newcommand{\partDeriv}[2]{\frac{\partial #1}{\partial #2}}  

\newcommand{\w}{\omega}
\newcommand{\dens}{m^{-3}}
\newcommand{\dg}{^\circ}   
\newcommand{\gn}{\grad n} 

\newcommand{\fref}[1]{Fig.~\ref{#1}}
\newcommand{\eref}[1]{Eq.~\ref{#1}}

\newcommand{\header}[1]{\section{#1}}

\newcommand{\ld}{lead}
\newcommand{\eq}{equal}
\newcommand{\su}{supporting}

\usepackage{ifthen}
\newcommand{\optCredit}[3][;]{
\ifthenelse{\equal{#3}{}}{}{#2 (#3)#1}}

\newcommand{\hlcolor}[2]{\sethlcolor{#1}\hl{#2}}
\newcommand{\hg}[1]{\hlcolor{green}{#1}}
\newcommand{\hy}[1]{\hlcolor{yellow}{#1}}
\newcommand{\hb}[1]{\hlcolor{cyan}{#1}}

\newcommand{\cbox}[2]{\fcolorbox{#1}{white}{#2}}
\newcommand{\gbox}[1]{\cbox{green}{#1}}

\if\countMode1
    \renewcommand{\cite}[1]{{}}
\fi

\if\hlMode0
    \renewcommand{\hlcolor}[2]{#2}
    \renewcommand{\cbox}[2]{#2}
\fi

\if\hlMode1
    \let\oldcite\cite
    \renewcommand{\cite}[1]{\mbox{\oldcite{#1}}}
\fi

\makeatletter
\def\@email#1#2{%
 \endgroup
 \patchcmd{\titleblock@produce}
  {\frontmatter@RRAPformat}
  {\frontmatter@RRAPformat{\produce@RRAP{*#1\href{mailto:#2}{#2}}}\frontmatter@RRAPformat}
  {}{}
}%
\makeatother

\begin{document}

\title{Preference of Right-Handed Whistler Modes and Helicon Discharge\\Directionality due to Plasma Density Gradients}
\author{M. Granetzny}
 \email{granetzny@wisc.edu}
\author{O. Schmitz}
\affiliation{University of Wisconsin - Madison, Department of Nuclear Engineering and Engineering Physics, WI, USA}
\author{M. Zepp}
\affiliation{University of Wisconsin - Madison, Department of Nuclear Engineering and Engineering Physics, WI, USA}

\date{\today}

\begin{abstract}
Whistlers are magnetized plasma waves in planetary magnetospheres. Bounded whistlers, known as helicons, can create high-density laboratory plasmas. We demonstrate reversal of the plasma discharge direction by changing either antenna helicity or magnetic field direction. Simulations reproduce these findings only in the presence of a radial density gradient. Inclusion of such a gradient in the wave equation gives rise to azimuthal shear currents which for the first time consistently explains the preference of right- over left-handed whistlers and the discharge directionality in helicon plasmas.\end{abstract}


\maketitle
\preprint{AIP/123-QED}


Helicons\cite{Boswell1984,Chen1991} were discovered in the 1960s: first in solids\cite{Aigrain1960,maxfield1969helicon}, then plasmas\cite{Lehane1965,Woods1962,Klozenberg1965}, and identified as bounded whistler waves{\cite{stenzel1999whistler}}. Whistlers are magnetized plasma waves and have been observed in the magnetospheres of Venus, Earth, Jupiter, Saturn, Uranus and Neptune\cite{zarka2004radio,strangeway1991plasma}. How these waves can be used in terrestrial applications to generate and probe plasmas has been of general interest and subject to several decades of intensive research. Laboratory helicons are radio-frequency (RF) waves, typically at tens of MHz, and are well known for their ability to excite steady-state high density plasmas up into the $\Eb{20}\U{\dens}$ range\cite{Buttenschon2018}, directly from a cold gas. This capability can be highly beneficial to multiple fields of science and engineering including semiconductor-etching\cite{Tynan1997}, advanced space propulsion\cite{Squire2006}, nuclear fusion materials testing\cite{Rapp2016,Rapp2017} and plasma wakefield acceleration\cite{Buttenschon2018,gschwendtner2016awake}. Their widespread occurrence in nature and the large range of applications makes understanding the confinement and directionality of these waves a high priority. Helicons are therefore of major interest inside the broader field of low-temperature plasma physics\cite{adamovich20222022,tsankov2022foundations}.

\begin{figure}
\includegraphics[width=\linewidth]{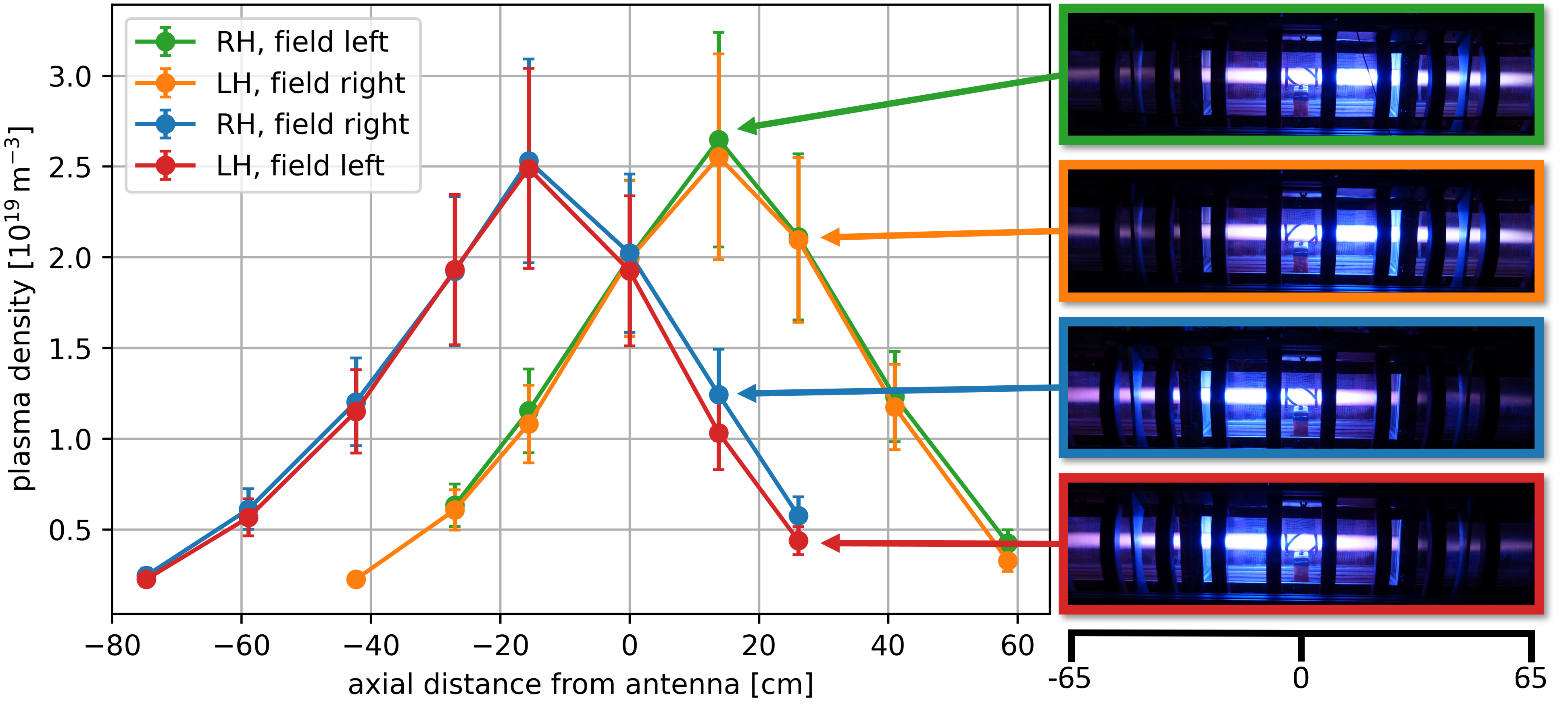}
\caption{\label{fig:directionalityExp}Dependence of the helicon plasma density and light emission on the antenna helicity - left-helical (LH) or right-helical (RH) - and background magnetic field direction - leftward or rightward in this figure.}
\end{figure}

\begin{figure}
\includegraphics[width=\linewidth]{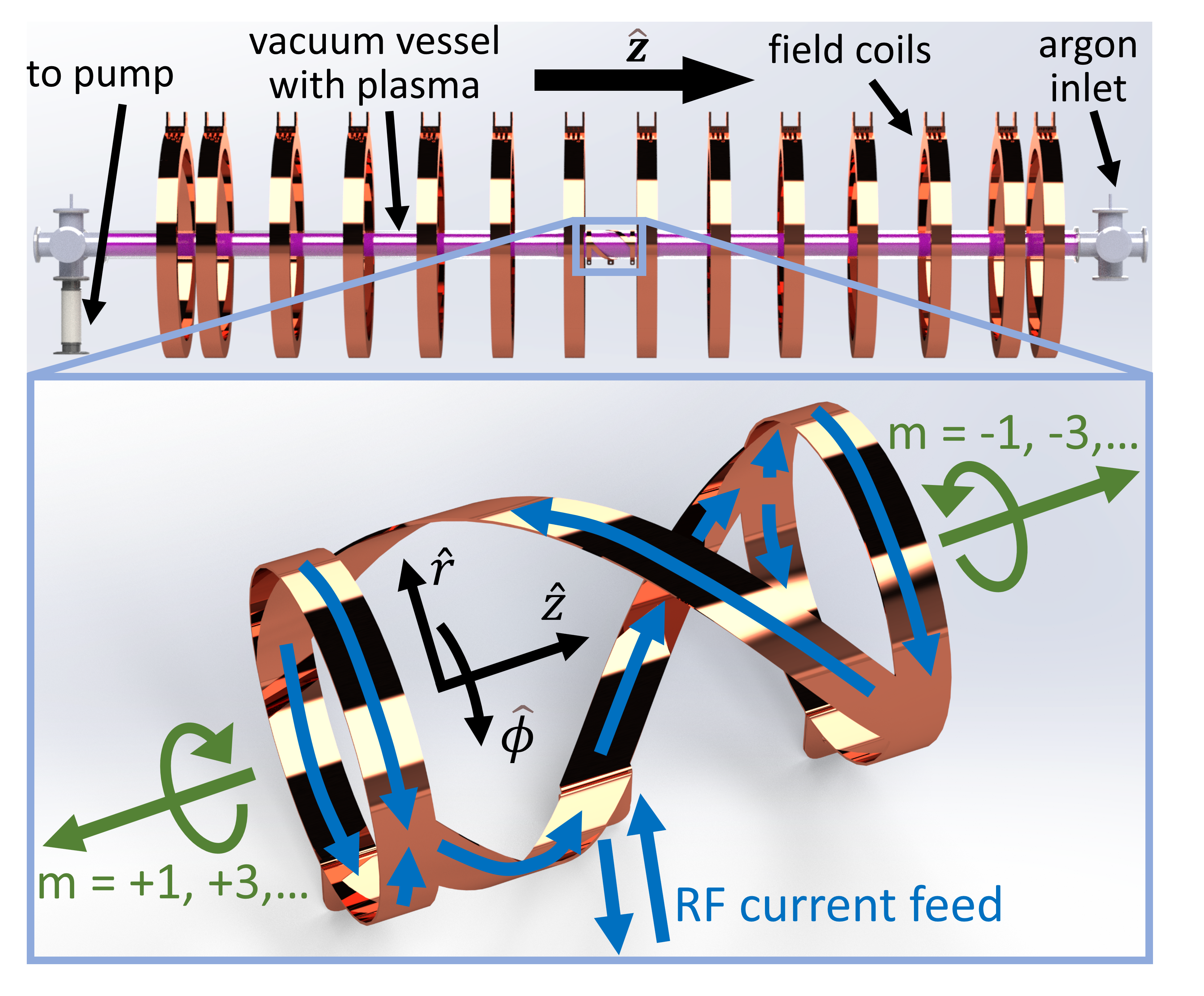}
\caption{\label{fig:Hardware}Top: CAD model of the core components of the MAP experiment. Bottom: Right-half-helical antenna with currents during one half of the RF cycle in blue and axial launch and azimuthal rotation directions for the different modes in green.}
\end{figure}

Whistler waves propagating along a magnetic field in an unbounded homogeneous plasma are R-waves{\cite{stix1992waves} and as such right-hand polarized. However, whistler modes in bounded plasmas, such as laboratory helicons or ducted magnetospheric whistlers\cite{karpman1982whistler,chen2021situ}, form mode patterns that can rotate in a left-handed or right-handed sense around the guiding magnetic field{\cite{Chen2015}}. Experiments have shown a predominance of right- over left-handed whistler modes{\cite{light1995helicon}}. Moreover, experiments using half-helical antennas{\cite{Miljak1998}}, which are known to produce the highest plasma density, have shown a strong directionality of the plasma discharge{\cite{Sudit1996}}. However, neither observation could ever be explained satisfactorily. We present here for the first time the reason for both effects and how they relate to each other.}
In {\fref{fig:directionalityExp}, the plasma directionality is demonstrated at the MAP experiment, shown later in {\fref{fig:Hardware}}. Density profiles and optical emission show that a right-helical antenna - a half-helical antenna with right-handed helicity - in a rightward magnetic field generates a leftward discharge (blue case). In a leftward field the discharge is rightward (green). For a left-helical antenna the discharge is to the right for a rightward field  (orange) but to the left for a leftward field (red). These measurements reveal that the discharge can be directed by changing the magnetic field direction or antenna helicity, resulting in identical but axially mirrored profiles. Starting from this observation, we demonstrate in this paper for the first time that radial density gradients enhance or attenuate whistler modes based on their handedness. For radially peaked density profiles this results in the preference of right-handed modes which in turn leads to the observed discharge directionality. This mechanism is fundamental to the coupling of whistler modes to magnetized plasmas with radial density gradients in solid-state, gaseous laboratory, and magnetospheric plasmas.}

Over the years numerous groups have studied half-helical antennas with right- and left-handed helicity  and found that the former produce higher plasma densities than the latter\cite{Sudit1996,Chen2015}. However, all those studies were performed in experiments with complicating effects due to the vacuum vessel or magnetic field geometry. The most common case is the excitation of helicon waves close to the axial boundary of the vacuum vessel\cite{Shinohara2004,Lehane1965,Sudit1996,Green2020,Thakur2015,Boivin2001,Miljak1998,Blackwell1997}, which introduces axially asymmetric boundary conditions around the antenna and can lead to discharges being launched directly into a nearby vacuum wall. In addition many experiments employ expansion chambers downstream of the helicon excitation region\cite{Degeling1996,Sung2016,Magee2013a,Chi1999,Balkey2001}, resulting in strong axial density gradients. Another common setup features strong gradients in the axial magnetic field\cite{Afsharmanesh2018,Piotrowicz2018,Sung2016,Guittienne2005,Balkey2001}. Inferring the reasons for the discharge directionality and preference of right-handed modes is difficult in these setups because the helicon dispersion relation depends strongly on the magnetic field strength and plasma density. To our knowledge no experimental investigation of the discharge directionality has been conducted in a setup without these complicating effect, thus the effect of field direction and antenna helicity on the discharge direction in {\fref{fig:directionalityExp}} has not been demonstrated before. For the experiments presented here we generated the helicon discharge from the center of a long vacuum vessel, with significant distance to both axial boundaries, and inside a highly homogeneous field.

The experiments were performed at the Madison AWAKE Prototype (MAP), shown in \fref{fig:Hardware}. MAP consists of a $2\U{m}$ long borosilicate glass tube with an ID of $52\U{mm}$ and an OD of $56\U{mm}$. $14$ coils produce a very homogeneous magnetic field of $49\U{mT}$ in the central $1.6\U{m}$ of the device, reaching $55\U{mT}$ at the ends. A $10$ cm long antenna with $1\U{cm}$ wide straps, shown in the lower panel of \fref{fig:Hardware}, is wound in either a right- or left-helical sense and used to excite the helicon plasma.  The right-half-helical antenna shown launches both right- and left-handed azimuthal modes, with mode numbers $m$, in the indicated directions\cite{Piotrowicz2018}. Notably, a whistler mode's handedness can be uniquely defined relative to the background magnetic field. In a field pointing along $\uniZ$ in \fref{fig:Hardware} positive $m$ modes are right-handed and negative $m$ modes are left-handed. Importantly, a left-helical antenna reverses the axial launch directions, making negative $m$ modes right-handed and positive $m$ modes left-handed.

All experiments were performed at an argon fill pressure of $\Eb{-2}\U{mbar}$ with RF input power set to $1.3\U{kW}$ at $13.56\U{MHz}$. An impedance matching network was used to reduce reflected power to negligible levels of less than $10\U{W}$. Plasma breakdown was readily achieved at $\w_{RF}/\w_{ce}=9.89\E{-3}$ and $B/n_n= 1.98\E{5} \U{Hx}${\cite{dujko2015heating}}, where $\w_{ce}$ and $n_n$ are the electron-cyclotron frequency and initial neutral gas density, respectively. The plasma density was measured by means of laser-induced fluorescence (LIF) on singly ionized argon. \hg{The LIF diagnostic is nearly identical to the LIF system used to study helicons in the MARIA device{\cite{Green2020,Green2020a}}. The diagnostic is designed around a 40 mW tunable diode laser, amplified to a maximum of 500 mW. The laser excites the argon ion transition at 668.614 nm from the $3$d$^4$F$_{7/2}$ state to the $4$p$^4$D$_{5/2}$ state, which then decays with an emission at 442.6 nm to the $4$s$^4$P$_{3/2}$ state within nanoseconds. The laser is linearly polarized and injected radially into MAP. Movable collection optics with a radial line of sight at a $90\dg$ angle to the injected beam collect the fluorescent light and deliver it to a photo-multiplier tube. Subsequently the fluorescence signal is extracted with a lock-in amplifier.}  In addition, photographs of the discharges were taken. 

The results were shown previously in \fref{fig:directionalityExp}. They show that reversal of either the antenna helicity or background field direction reverses the discharge direction and reversal of both restores the original discharge direction. \hg{The matching network settings used were identical for all four cases and resulted in less than $10 \U{W}$ reflected power for each of them. This observation indicates identical plasma impedance across all four discharges.} These results were obtained with very good magnetic field homogeneity and proper distance to axial vacuum boundaries. They therefore remove ambiguities in the interpretation of measurements by other groups due to their experimental setups, such as placement of the antenna close to an axial boundary or near regions with plasma density or field strength outside the values allowed by the dispersion relation\cite{Chen1997}.

Helicon discharges in MAP were modelled using a quasi-3D finite element model developed in COMSOL using the cold plasma wave description. In a high density helicon plasma a significant part of the power is deposited by the Trivelpiece-Gould mode\cite{Arnush2000a} which has very short radial wavelength on the sub-mm scale. This necessitates use of very fine mesh elements of $500 \U{\mu m}$ axially and $30\U{\mu m}$ radially. The model assumes that wavefields have an $e^{im\phi}$ dependence in the azimuthal direction. Due to strong damping of higher order modes, a full 3D solution can be calculated from the six leading order azimuthal modes, namely $m = \{\pm1, \pm3, \pm5\}$. Power deposition is calculated strictly ohmically through electron-ion and electron-neutral collisions since Landau damping is negligible\cite{Chen1999UpperDischarges}. The model was verified by comparing results in a homogeneous plasma against the analytical helicon dispersion relation\cite{Chen1997}.

\begin{figure}
\includegraphics[width=\linewidth]{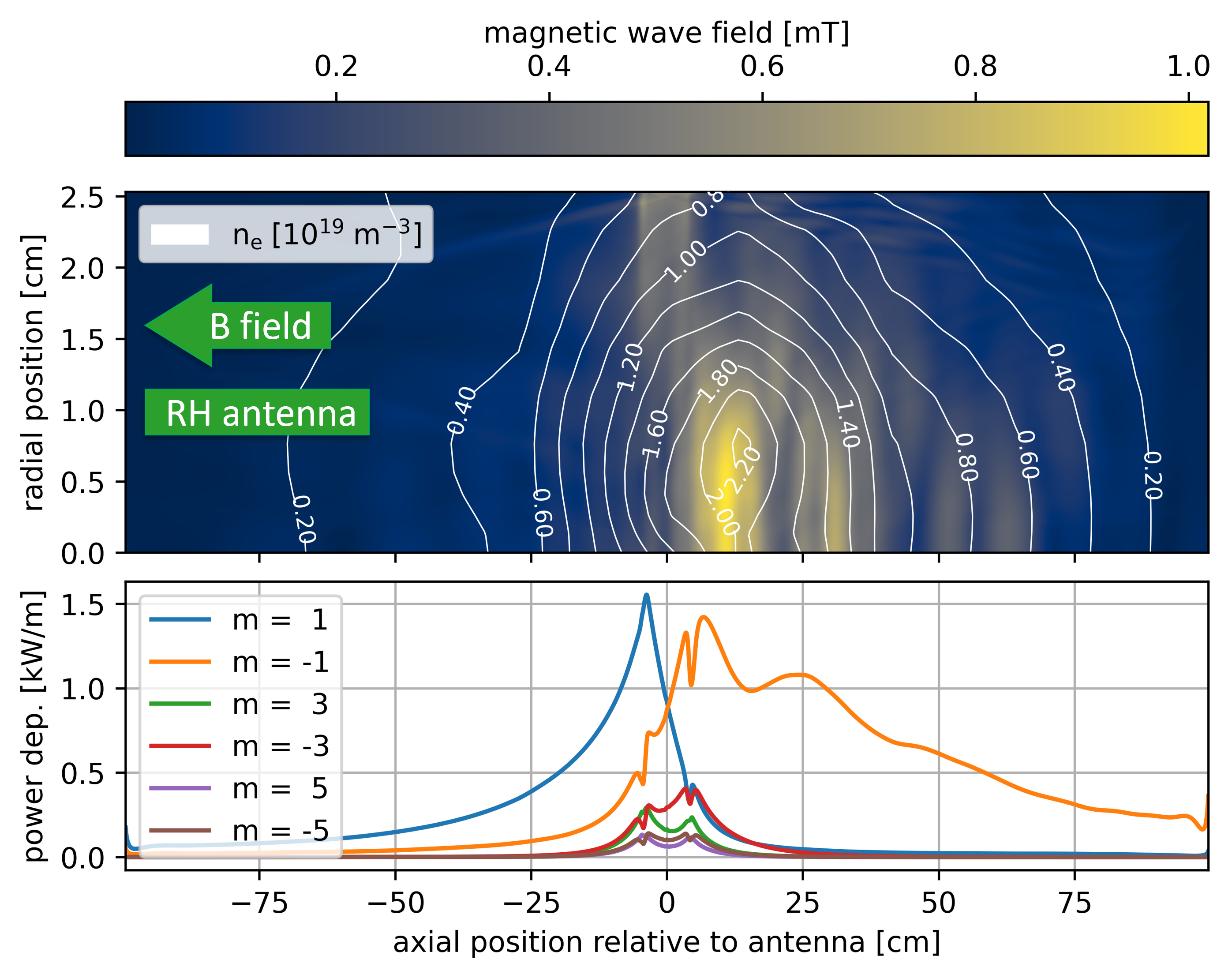}
\caption{\label{fig:wavefieldDens}Simulation results for an RH antenna with magnetic field in the negative $\uniZ$ direction, representing the green case in {\fref{fig:directionalityExp}}. Top: Measured plasma density and simulated wave magnetic field strength. Bottom: Axial power deposition by azimuthal modes. }
\end{figure}

Plasma temperature and neutral pressure were set to be uniform at $3\U{eV}$ and $\Eb{-3}\U{mbar}$, respectively, the latter assuming a $90$\% neutral depletion. Density profiles used in the simulation were interpolated from \hy{$86$} local LIF measurements in a plasma generated by a right-helical antenna in a leftward field (green case in \fref{fig:directionalityExp}). The results of this simulation are shown in \fref{fig:wavefieldDens}. The upper panel  shows the magnitude of the total wave magnetic field, with plasma density indicated by white contour lines. The helicon wave propagates radially inwards and axially rightward from the antenna location. The lower panel shows the radially integrated power deposition contributions from the different azimuthal modes. \hg{Azimuthal modes with negative mode numbers (-1, -3, ...) deposit power predominantly to the right, while positive modes (1, 3, ...) deposit power mostly to the left. This is in agreement with the azimuthal mode launch directions for these modes as indicated in {\fref{fig:Hardware}}.} In this setup the dominant right-handed mode, $m=-1$, deposits $58.7$\% of the total power, mostly to the right, while the leading left-handed mode, $m=+1$, deposits $29.5$\% of the total power, mostly to the left. The remaining $11.8$\% are accounted for by higher order modes. Overall the power deposition is dominated by the leading right-handed mode and shows the same directionality observed optically and by LIF in \fref{fig:directionalityExp} (green case). 

\begin{figure}
\gbox{
\includegraphics[width=\linewidth]{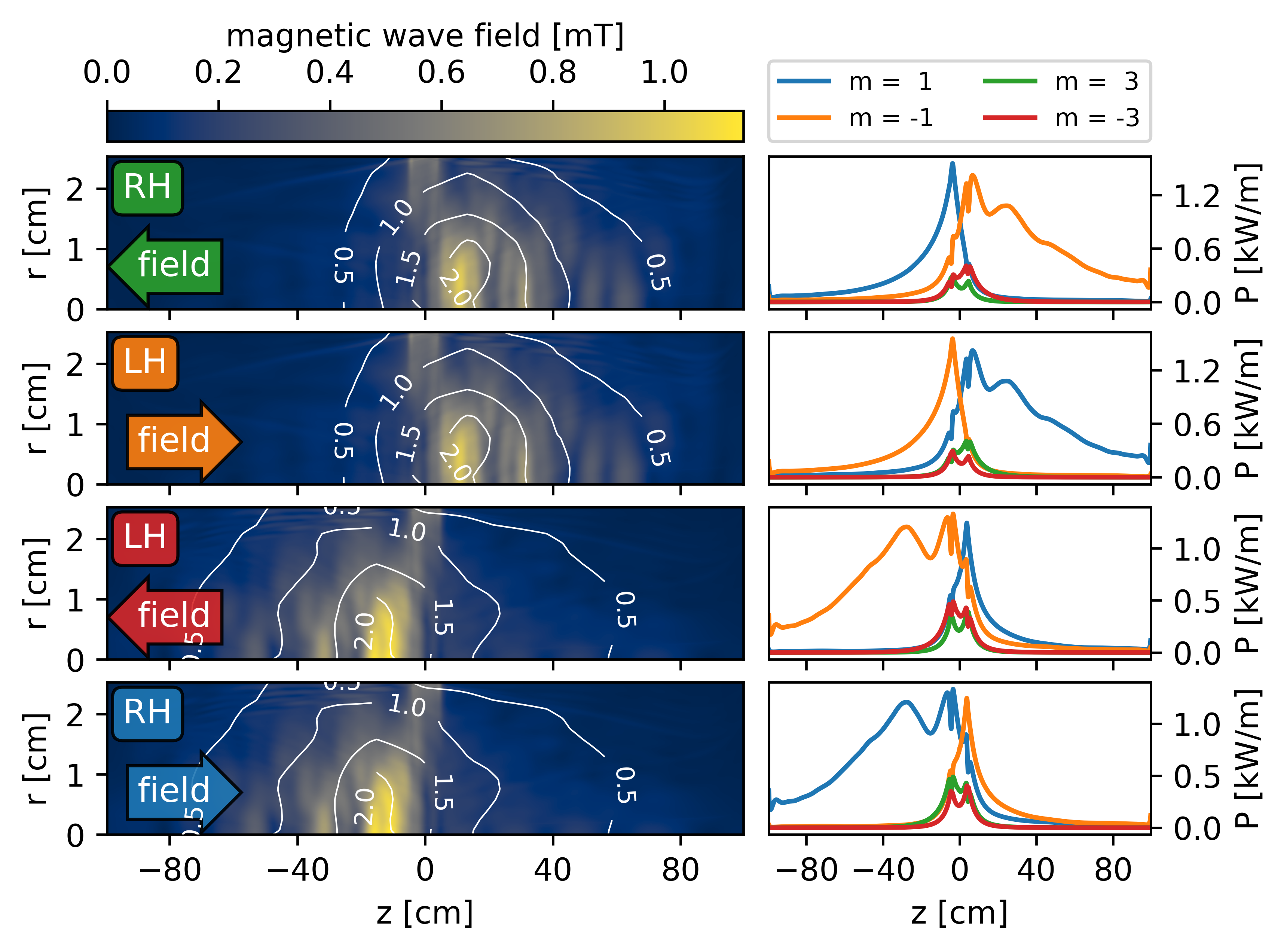}}
\caption{\label{fig:LIFSimComp}\hg{Comparison of simulation results for the four combinations of antenna helicity and magnetic background field direction, representing the green, orange, red and blue cases in {\fref{fig:directionalityExp}}. White contour lines show plasma density in units of $\Eb{19}\U{\dens}$}.}
\end{figure}

\hg{A comparison of simulation results for all four combinations of antenna helicity and background field direction is shown in {\fref{fig:LIFSimComp}}. The density profiles shown were measured for a right-handed antenna with both leftward and rightward pointing background fields. Due to the symmetry shown experimentally in {\fref{fig:directionalityExp}}, these density profiles were used for simulating the wave propagation from left-handed antennas as well. As expected from our experiments these simulations show a reversal of the wavefield propagation direction and power deposition patterns if either the background field or antenna helicity is reversed. The power deposition is dominated by either the $m=1$ or $m=-1$ azimuthal modes, depending on which is the right-handed mode with respect to the background field direction. As mentioned earlier reversal of the antenna helicity reverses the launch directions for these azimuthal modes.}

To shed light on the reason for the directional power deposition we performed simulations with axially uniform plasma density profiles. The radial plasma density profiles were either modelled as a flat top of the form\cite{Arnush1998}
\begin{align}
    \label{eq:profileShape}
    n_e &= \br{n_e^{peak}-n_e^{edge}}\sbr{1-\left(\frac{r}{r_w}\right)^5}^5 + n_e^{edge},
\end{align}
\noindent
with $n_e^{peak} = 2.5\E{19}\U{\dens}$, $n_e^{edge} = 5.0\E{18}\U{\dens}$ and vacuum wall location $r_w = 26 \U{mm}$ or with a constant density of $n_e = 2.5\E{19}\U{\dens}$ throughout.

\begin{figure}
\includegraphics[width=\linewidth]{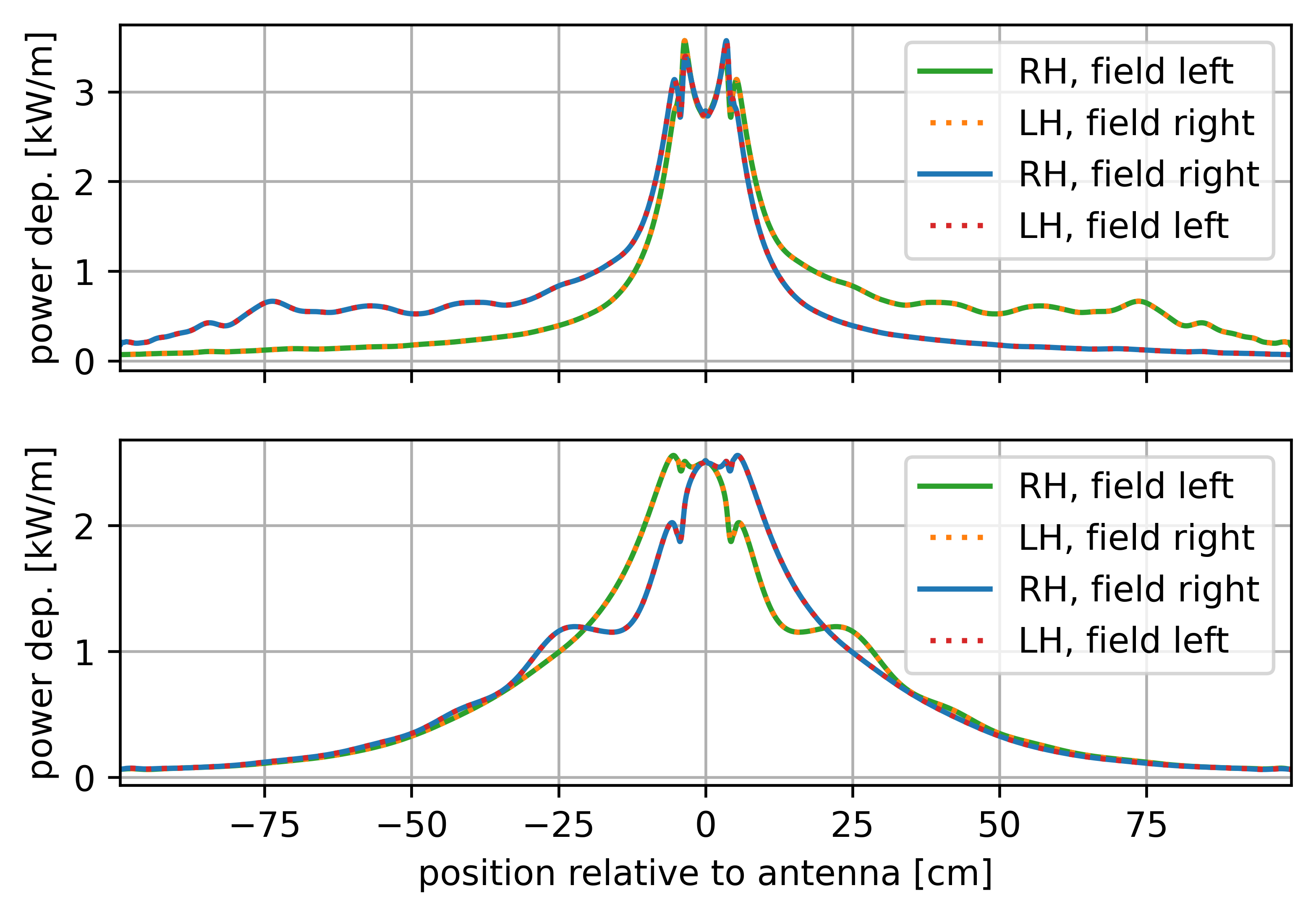}\caption{\label{fig:powerDepSim}Simulated power deposition profiles for helicon discharges with different field directions and antenna helicities in plasmas with radial density gradient (top) and uniform density (bottom). Green and orange as well as blue and red curves are overlapping, respectively.}
\end{figure}

Results for the four combinations of antenna helicity and field direction for both profiles are shown in \fref{fig:powerDepSim}. For a radial density gradient and a rightward field with an right-helical antenna, or a leftward field with an left-helical antenna, the power deposition is predominantly to the left. For the other two combinations, the power deposition is predominantly to the right. The left-right power imbalance is 36.5\% to 63.5\%. For the radially homogeneous plasma no significant preferential power deposition exists, with the left-right imbalance being only 48.1\% to 51.9\%.

Measurements and simulations show qualitatively the same effect of antenna helicity and magnetic field direction on the discharge direction. We have shown computationally that, even in an axially homogeneous plasma, a radial density gradient leads to a preferential direction for axial power deposition. However, the power deposition becomes symmetric when no radial density gradient is present. An understanding of this phenomenon can be gained by deriving the helicon wave equation for a plasma with a purely radial density gradient. We follow Chen's approach\cite{Chen1997} who derives the wave equation for a completely homogeneous plasma. Wave fields are assumed to be sinusoidal in the azimuthal and axial directions, i.e. they are of the form $e^{i\left(m\phi + k z - \w t\right)}$, where $k$ is the axial wavenumber and $\w$ the wave frequency. In helicon plasmas the electron currents are far stronger than the displacement currents and the ions are immobile, hence the relevant Maxwell equations in the frequency domain become
\begin{eqnarray}
\curl \vE &&= i \w \vB\label{eq:FaradaysLaw}\\
\curl \vB &&\approx \mu_0 \vj \approx - \mu_0 e n \vv, \label{eq:AmperesLaw}   
\end{eqnarray}
\noindent
where the incompressible electron fluid has density $n$ and velocity $\vv$. The electrons are subject to electric, magnetic, friction and pressure gradient forces such that their momentum equation becomes
\begin{equation}
- i \w m_e \vv = - e \br{ \vE + \vv \times \vec{B_0}} - m_e \nu \vv - \frac{k_b T_e}{n}\grad n \label{eq:eleMom},
\end{equation}
\noindent
where $\vec{B_0}$ is the background magnetic field, $T_e$ is the electron temperature and $\nu$ is an effective combined electron-ion and electron-neutral collision frequency.

As shown in the appendix, we can use this set of equations to eliminate $\vE$, $\vj$ and $\vv$ and arrive for purely radial density gradients at
\begin{eqnarray}
&&\delta \curl\curl \vB \mp k \curl \vB + k_w^2 \vB \nonumber\\
&&=  \delta \frac{\grad n}{n} \times (\curl \vB) \mp i \sbr{\frac{\grad n}{n} \cdot (\curl \vB)}\uniZ\label{eq:waveTwo},
\end{eqnarray}
\noindent
where we have used the definitions
\begin{equation}
\label{eq:deltaDef}
\delta = \frac{(\w + i \nu)m_e}{e B_0}\quad\text{and}
\quad k_w^2 = \frac{\mu_0 e n \w}{B_0}.
\end{equation}
In \eref{eq:waveTwo} the $\mp$ sign accounts for pointing of the background magnetic field either along or against $\uniZ$. The left-hand side of \eref{eq:waveTwo} is the helicon wave equation in a homogeneous plasma\cite{Chen1997}. The terms on the right-hand side exist only in the presence of a density gradient; we will hereafter refer to them as the modulating magnetic field as explained below. Since $\delta$ is the ratio of the RF frequency to the electron cyclotron frequency, the first  term on the right-hand side of \eref{eq:waveTwo} is negligible and the modulating magnetic field becomes
\begin{equation}   
\mp i \sbr{\frac{\grad n}{n} \cdot (\curl \vB)}\uniZ \approx \pm \frac{m B_z}{nr}\frac{\partial n}{\partial r}\uniZ,\label{eq:waveFour}
\end{equation}
\noindent
where the right-hand side exploits that $r$ becomes small towards the plasma core where the density is highest such that the $B_z$ term of $\curl \vB$ dominates.

This form shows very clearly a dependence on the magnetic field direction, density gradient and azimuthal mode number. For a background field along $\uniZ$ and a radially inward density gradient, the additional modulating magnetic field enhances positive but attenuates negative $m$ modes, which explains the preference of right-handed over left-handed modes. Moreover, since helical antennas of opposite helicity send those modes in opposite directions this explains the directionality of helicon discharges and why helicity or field reversal flip the discharge direction. \hb{This mechanism also explains the remaining power imbalance in the homogeneous plasma case in {\fref{fig:powerDepSim}}. In our model the antenna launches waves at the outside of the vacuum vessel which has a wall thickness of 2 mm. This represents an effective very steep radial plasma density gradient from the launch region into the plasma edge. The result is a very strong attenuation of left-handed modes in this transition region which leads to a high power deposition close to the antenna. As the waves propagate radially further inward and axially away from the launch region the power deposition becomes symmetric again as the radial plasma density gradient is zero in the homogeneous plasma case.}

The significance of the modulating magnetic field relative to the wave dynamics in a uniform plasma becomes clear by comparing the $k \curl \vB$ term in \eref{eq:waveTwo} to the dominant source term $\frac{\grad n}{n} \cdot (\curl \vB)$. Their ratio, i.e. $\frac{\grad n}{n}/k \approx (rk)^{-1}$, is 63\%, where we have used the vessel radius $r=26 \U{mm}$ and an axial wavenumber $k=2\pi/(10 \U{cm})$ defined by the antenna length. The density gradient thus results in a significant change in the wave dynamics. \hg{Importantly the derivation of the wave equation makes no assumption on the strength of the background magnetic field apart from that it should result in the whistler wave frequency being much smaller than the electron-cyclotron frequency such that we have $\delta \ll 1$. For example at our RF frequency of 13.56 MHz this assumption would still be valid down to a field of just $5 \U{mT}$ resulting in a ratio of 0.1.} 

The physical mechanism behind \hy{modulating magnetic field} can be understood by using \eref{eq:AmperesLaw} to bring the wave equation into the form
\begin{equation}
\delta \curl\curl \vB \mp k \curl \vB + k_w^2 \vB = \mp \frac{i \mu_0}{n} \frac{\partial n}{\partial r} j_r \uniZ,\label{eq:waveThree}    
\end{equation}
\noindent
which shows that radial wave currents ($j_r$) in conjunction with the field direction and density gradient are responsible for the additional modulating magnetic field which has a $\mp90\dg$ phase shift relative to $j_r$.

\begin{figure}
    \includegraphics[width=\linewidth]{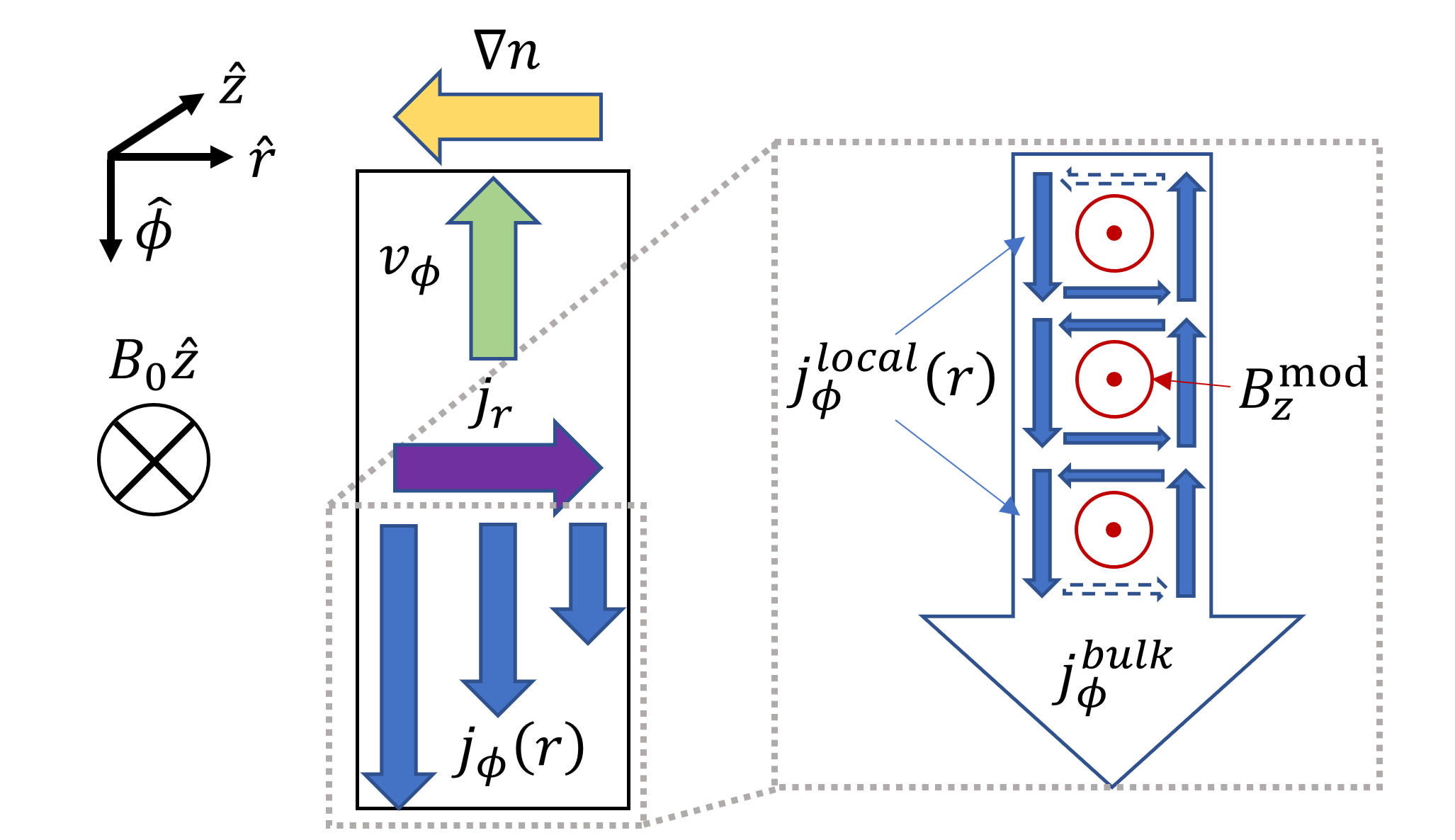}\caption{\label{fig:reinforcementMech}Mechanism behind the modulating magnetic field in the presence of a radial density gradient.}
\end{figure}

A physical explanation of the modulating field is given in \fref{fig:reinforcementMech}. In a background field along $\uniZ$ an electron fluid element carrying a radial current $j_r$ is subject to a Lorentz force that induces a fluid flow ($v_\phi$) which in turn represents an azimuthal current $j_\phi$. This current experiences a Lorentz force as well, resulting in a current in the negative radial direction. This is the mechanism by which  helicon waves exchange energy between radial and azimuthal currents in a homogeneous plasma. However, in the presence of an electron density gradient ($\gn$) neighboring fluid elements will carry azimuthal currents with magnitudes proportional to the electron density, leading to an azimuthal shear current. These currents can be described by a bulk current plus local currents for the different radial positions. The local parts represent dipole currents which create a magnetic field in the $\pm\uniZ$ direction which we have expressed mathematically by the modulating magnetic field in \eref{eq:waveThree}. For a right-handed helicon mode in a homogeneous plasma the $j_r$ currents lead $j_\phi$ by $90\dg$. In the case of a radially inward density gradient, the modulating field then points along the mode's regular $B_z$ field and leads to an enhancement of the wavefields. In contrast a left-handed mode has $j_r$ lagging $j_\phi$ by $90\dg$ such that the additional modulating field results in $B_z$ wavefields in the wrong direction, thereby attenuating $j_r$, $j_\phi$ and the wave overall. In a hollow plasma channel, such as density troughs guiding magnetospheric whistler waves{\cite{stenzel1999whistler}}, the same mechanism would lead to an enhancement of left- over right-handed modes.

\begin{figure}[tbh]
\includegraphics[width=\linewidth]{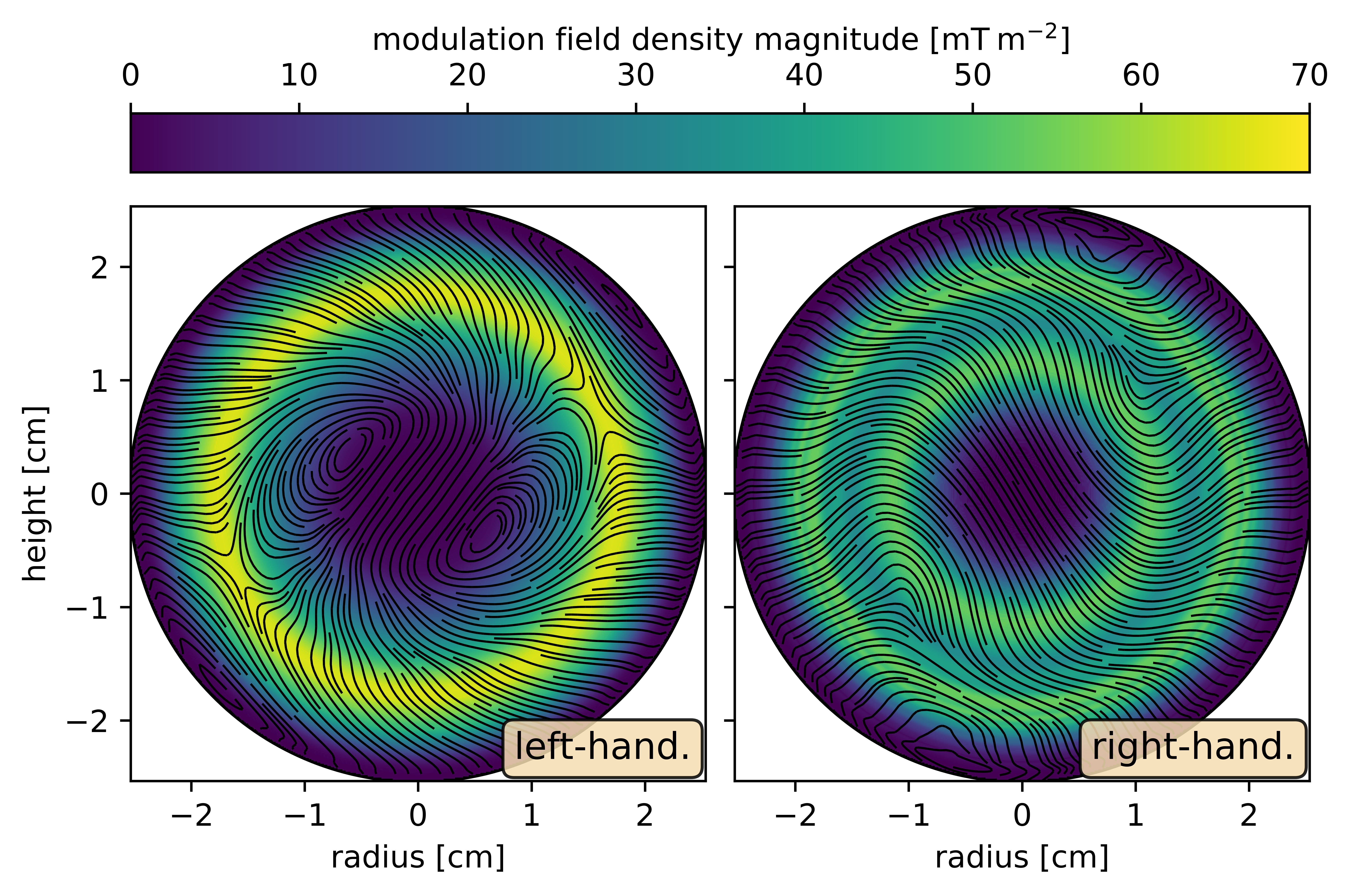}
\caption{\label{fig:fieldLineComp}Magnitude of the modulating magnetic field and wave magnetic field lines for the leading left-handed and right-handed modes.}
\end{figure}

The modulating magnetic field results in a distortion of the left- and right-handed mode patterns compared to those in a homogeneous helicon plasma. This is demonstrated in \fref{fig:fieldLineComp} which shows cross sections of the leading left- and right-handed mode magnetic wave fields superimposed on the magnitude of the modulating field. The enhancement of the right-handed mode is visible by a lack of vortices that are normally present in such modes\cite{Chen1996a} in addition to a general dragging of the field lines in the direction of rotation for this mode, i.e. in this view counterclockwise. For the left-handed mode the strength of vortices is increased and we find regions of field reversal at the radial locations with strong modulation fields. Field lines are dragged counterclockwise as well which is against the direction of rotation for this mode.

We have shown experimentally, computationally and analytically that radially inward density gradients enhance right- but attenuate left-handed whistler modes. This mechanism explains long standing observations of preferential excitation of right-handed whistler modes and consequently the directionality of helicon plasmas generated by helical antennas. The discharge direction is defined by the combination of antenna helicity and magnetic background field direction. Measurements show that all four combinations of helicity and field direction produce identical discharges with the only difference being an axial mirroring around the antenna location. We have demonstrated that radial density gradients induce azimuthal shear currents which in turn generate a modulating magnetic field that interferes with the axial component of the whistler wave fields. The opposite phasing of currents in right- and left-handed whistler modes results in the modulating field amplifying the former but attenuating the latter. Since helical antennas of opposite helicity send right- and left-handed modes in opposite directions, changing the helicity changes the discharge direction. Moreover, since right- or left-handedness of a mode with a given spatial rotation direction depends on the direction of the background field, the discharge is reversed when the field direction is reversed. In a broader context, the discovered mechanism predicts preference of left-handed modes in hollow plasma channels and is fundamental to the wave-plasma coupling of whistler modes in general, for example in solids and planetary magnetospheres.

We would like to express our gratitude to Barret Elward for his outstanding work during the construction of the MAP experiment. We would also like to thank Dieter Boeyaert, Kelly Garcia, Madelynn Knilans, Carl Sovinec and Danah Velez for their feedback on the manuscript. The research presented here was funded by the National Science Foundation under grant PHY-1903316 and NSF-CAREER award PHY-1455210 as well as the College of Engineering at UW-Madison.

\section*{Author Declarations}
\subsection*{Conflict of Interest}
The authors have no conflicts to disclose.
\subsection*{Author Contributions}
\textbf{Marcel Granetzny: }
\optCredit{Conceptualization}{\ld}
\optCredit{Data curation}{\ld}
\optCredit{Formal analysis}{\ld}
\optCredit{Funding acquisition}{}
\optCredit{Investigation}{\ld}
\optCredit{Methodology}{\ld}
\optCredit{Project administration}{\eq}
\optCredit{Resources}{}
\optCredit{Software}{\ld}
\optCredit{Supervision}{}
\optCredit{Validation}{\eq}
\optCredit{Visualization}{\ld}
\optCredit{Writing - original draft}{\ld}
\optCredit[.]{Writing - review \& editing}{\ld}
\textbf{Oliver Schmitz: }
\optCredit{Conceptualization}{\eq}
\optCredit{Data curation}{}
\optCredit{Formal analysis}{}
\optCredit{Funding acquisition}{\ld}
\optCredit{Investigation}{}
\optCredit{Methodology}{}
\optCredit{Project administration}{\eq}
\optCredit{Resources}{\ld}
\optCredit{Software}{}
\optCredit{Supervision}{\ld}
\optCredit{Validation}{}
\optCredit{Visualization}{}
\optCredit{Writing - original draft}{\su}
\optCredit[.]{Writing - review \& editing}{\su}
\textbf{Michael Zepp: }
\optCredit{Conceptualization}{}
\optCredit{Data curation}{\eq}
\optCredit{Formal analysis}{\eq}
\optCredit{Funding acquisition}{}
\optCredit{Investigation}{\eq}
\optCredit{Methodology}{\eq}
\optCredit{Project administration}{}
\optCredit{Resources}{}
\optCredit{Software}{\eq}
\optCredit{Supervision}{}
\optCredit{Validation}{\eq}
\optCredit{Visualization}{}
\optCredit{Writing - original draft}{}
\optCredit[.]{Writing - review \& editing}{\su}

\section*{Data Availability}
The data that support the findings of this study are available from the corresponding author upon reasonable request.

\appendix
\setcounter{equation}{0}
\renewcommand{\theequation}{A\arabic{equation}}

\header{Derivation of the Wave Equation in a Plasma with Radial Density Gradient}
The plasma is assumed to be inhomogeneous only in the $\uniR$ direction such that any wavefields $\vec{W}$ are of the form

\begin{equation}
\vec{W} = \vec{W_0}(r)e^{i(m\phi + kz -\w t)}\label{eq:app0}.
\end{equation}

\noindent
Azimuthal and axial derivatives then simplify to

\begin{equation}
\partDeriv{\vec{W}}{\phi} = i m \vec{W},\label{eq:app0b}
\end{equation}

and,

\begin{equation}
\partDeriv{\vec{W}}{z} = i k \vec{W}.\label{eq:app0c}
\end{equation}

We will express the axial background magnetic field as
\begin{equation}
\vec{B_0} = sB_0\uniZ\quad \text{with}\quad B_0 = |\vec{B_0}|\label{BB},
\end{equation}

\noindent
where $s=\pm$ indicates alignment along or against $\uniZ$, respectively.\\

Using this expression in the electron momentum equation (\eref{eq:eleMom}) we get after some rearranging
\begin{equation}
\vE = B_0\sbr{s \crp{\uniZ}{\vv} + i \frac{\br{\w + i \nu}m_e}{e B_0}\vv} - \frac{k_b T_e}{e n}\gn.
\end{equation}

We can now substitute $\vE$ into Faraday's law (\eref{eq:FaradaysLaw}) and define $\delta = (\w + i \nu)m_e/(e B_0)$ to arrive at

\begin{equation}
i \w \vB = B_0\curl \br{s \crp{\uniZ}{\vv} + i \delta\vv}- \frac{k_b T_e}{e}\curl \frac{\gn}{n}.\label{eq:app1b}    
\end{equation}

By the chain rule we have

\begin{equation}
\grad \br{\frac{1}{n}} = -\frac{1}{n^2} \grad n,\label{gradInvN}
\end{equation}

\noindent
and the last term in \eref{eq:app1b} vanishes since

\begin{eqnarray}
    \curl \frac{\gn}{n} = \frac{\curl \gn}{n} - \frac{\crp{\gn}{\gn}}{n^2} = 0.
\end{eqnarray}

Further, by expressing $\vv$ through $\vB$ using \eref{eq:AmperesLaw}, we can transform \eref{eq:app1b} into

\begin{equation}
i \w \vB =-\frac{B_0}{\mu_0 e}\curl \pbr{\frac{1}{n}\sbr{s \crp{\uniZ}{\bcurlB} + i \delta\curlB}}.\label{eq:app1}   
\end{equation}

After applying the chain rule and \eref{gradInvN}, \eref{eq:app1} becomes

\begin{eqnarray}    
\frac{i \mu_0 e n \w}{B_0}\vB =&& - s \curl \sbr{\crp{\uniZ}{\bcurlB}} \nonumber\\
    &&- i \delta\curl \bcurlB\nonumber\\
    &&+\frac{s}{n} \crp{\grad n}{\sbr{\crp{\uniZ}{\bcurlB}}} \nonumber\\&&+ \frac{i \delta}{n} \crp{\grad n} {\bcurlB} \label{eq:app2}.
\end{eqnarray}  

By using the vector identity for the curl of a cross product, we can expand the first term in \eref{eq:app2} into four parts, namely

\begin{eqnarray}
    \bcurl{\crp{\uniZ}{\bcurlB}} =&& \uniZ {\sbr{\div \bcurlB}} \nonumber\\
    &&- \bcurlB {\br{\div\uniZ}}\nonumber\\
    &&+ \matDerivSq{\bcurlB}\uniZ \nonumber\\
    &&- \matDeriv{\uniZ}\bcurlB.\label{eq:app5}
\end{eqnarray}    

Since the divergence of a curl is zero and $\uniZ$ is constant, the first three terms in \eref{eq:app5} vanish. The last term can be calculated as

\begin{eqnarray}
- \matDeriv{\uniZ}\bcurlB =&& - \mu_0 \br{\uniR\partDeriv{j_r}{z}+\uniPhi\partDeriv{j_\phi}{z}+\uniZ\partDeriv{j_z}{z}}\nonumber\\
=&&- ik \bcurlB,
\end{eqnarray}

\noindent where we have exploited that $\curlB = \mu_0 \vj$ is of the form in \eref{eq:app0} such that \eref{eq:app0c} applies. Applying triple product expansion to the third term in \eref{eq:app2} yields

\begin{eqnarray}
\bcrpSq{\gn}{\bcrp{\uniZ}{\curlB}} =&& \uniZ \sbr{\dtp{\gn}{\bcurlB}}\nonumber\\
&&- \bcurlB{\br{\dtp{\gn}{\uniZ}}},\nonumber\\
\end{eqnarray}

\noindent
where the second term vanishes since $\gn \perp \uniZ$. \eref{eq:app2} then becomes

\begin{eqnarray}
i k_w^2\vB =&& iks\bcurlB - i \delta\curl \bcurlB\nonumber\\ 
    &&+ \frac{s}{n} \sbr{\dtp{\gn}{\bcurlB}}\uniZ + \frac{i \delta}{n} \crp{\grad n} {\bcurlB}\label{eq:app3},\nonumber\\
\end{eqnarray}

\noindent
where we have used the definition $k_w^2 = \mu_0 e n \w/B_0$. Lastly \eref{eq:app3} can be recast into

\begin{eqnarray}
&&\delta\curl \bcurlB -ks\bcurlB + k_w^2\vB \nonumber\\
&&= - i s\uniZ \sbr{\dtp{\frac{\gn}{n}}{\bcurlB}} + \delta \crp{\frac{\grad n}{n}} {\bcurlB},\nonumber\\    
\end{eqnarray}

which after back-substituting $s=\pm$ is the sought after wave equation shown in \eref{eq:waveTwo}. \hg{Since $\delta$ is of order $\Eb{-2}$ or smaller - e.g. $9.7\E{-3}$ at 13.56 MHz and 50 mT - the second source term is negligible.} Further, since $\gn \| \uniR$, by means of \eref{eq:app0b} and \ref{eq:app0c} the source simplifies into the alternate form

\begin{equation}
\mp i \sbr{\dtp{\frac{\gn}{n}}{\bcurlB}}\uniZ = \pm \frac{1}{n}\partDeriv{n}{r} \br{\frac{m B_z}{r} - k B_\phi}\uniZ,
\end{equation}

\noindent
which for comparable strength of $B_z$ and $B_\phi$ and $r^{-1} \gg k$ becomes the expression shown in \eref{eq:waveFour}. Alternatively, since $\curlB = \mu_0 \vj$, the source for a purely radial density gradient becomes

\begin{equation}
\mp i \sbr{\dtp{\frac{\gn}{n}}{\bcurlB}}\uniZ = \mp \frac{i \mu_0}{n} \frac{\partial n}{\partial r} j_r \uniZ,    
\end{equation}

\noindent
as shown previously in \eref{eq:waveThree}.


\header{\hg{Symbols used in this Work}}
\begin{table}[h]
\begin{ruledtabular}
\begin{tabular}{ll}
Symbol & Description\\
\hline
 $m$ & azimuthal mode/wave number  \\ 
 $k$ & axial wave number \\
 $\omega$ & whistler/helicon wave frequency [rad/s]  \\ 
 $\vE$ &  electric wavefield components \\ 
 $\vB$ &   magnetic wavefield components\\ 
 $\vj$ &   wavefield currents\\ 
 $B_0$ & background magnetic field magnitude \\
 $s$ & direction of magnetic background field (\eref{BB}) \\
 $n$ & electron density  \\ 
 $\vv$ & electron velocity  \\ 
 $T_e$ & electron temperature  \\ 
 $\nu$ & effective combined electron-ion\\
 & and electron-neutral collision frequency\\ 
 $\delta$ & ratio of whistler to electron-cyclotron frequency (\eref{eq:deltaDef})\\
 $k_w$ & $\sqrt{\mu_0 e n \w /B_0}$ (\eref{eq:deltaDef})\\ 
 \end{tabular}
\end{ruledtabular}
\end{table}
\section*{References}
\nocite{*}
\bibliography{references}\textbf{}
\end{document}